# Tunneling transport of mono- and few-layers magnetic van der Waals MnPS$_3$

Sungmin Lee,[1,2] Ki-Young Choi,[1] Sangik Lee,[3] Bae Ho Park,[3] and Je-Geun Park[1,2,a]

[1]*Center for Correlated Electron Systems, Institute for Basic Science, Seoul 08826, Korea*

[2]*Department of Physics & Astronomy, Seoul National University, Seoul 08826, Korea*

[3]*Department of Physics, Konkuk University, Seoul 05029, Korea*

We have investigated the tunneling transport of mono- and few-layers of MnPS$_3$ by using conductive atomic force microscopy. Due to the band alignment of indium tin oxide/MnPS$_3$/Pt-Ir tip junction, the key features of both Schottky junction and Fowler-Nordheim tunneling (FNT) were observed for all the samples with varying thickness. Using the FNT model and assuming the effective electron mass (0.5 $m_e$) of MnPS$_3$, we estimate the tunneling barrier height to be 1.31 eV and the dielectric breakdown strength as 5.41 MV/cm.

The discovery of graphene and its rich physics subsequently unearthed have literally revolutionized the field of van der Waals (vdW) materials and the list of newly found vdW systems are exploding at a fast pace[1-3]. However, it became clear soon after the initial discovery that although the zero band gap of graphene renders the associated physic fascinating, at the same time it poses serious problems for potential applications using graphene. This realization has driven more recent searches for vdW materials with a sizeable band gap such as hexagonal boron nitride and transition metal di-chalcogenides (TMDCs).[4]

Upon close inspection of the ever growing list of the vdW materials, however one can take an immediate note of the fact that there is hardly any magnetic vdW system in the list, with the only exception of the High-$T_c$ superconductor $Bi_2Sr_2CaCu_2O_{8+d}$.[5,6] Without citing numerous contributions magnetic materials have made over past many decades, if not centuries, to the fundamental physics as well as the industrial applications, the magnetic vdW systems, once discovered, will form an invaluable part of applications in fields like spintronics.[7] And then, there is huge potentials yet untapped when it is used in tandem with other vdW materials as components of vdW heterostructures.[8]

[a] Electronic mail: jgpark10@snu.ac.kr

Thus it is an interesting and welcome development to see the recent entry of new magnetic vdW materials to the field in the name of transition metal phosphorus tri-chalcogenide ($TMPX_3$). One of the big advantages with $TMPS_3$ is that it can host several transition metal elements at the TM sites with correspondingly diverse physical properties: TM = Mn, Fe, Co, and Ni. A passing note, nonmagnetic Zn and Cd at the TM site also form in the same phase. Different TM elements lead to different magnetic ground states and varying transition temperatures although all of them exhibit an antiferromagnetic ordering of one type or another with the transition temperature around 80 - 150 K.[9] For instance, $MnPS_3$ has an antiferromagnetic order of type II below $T_N$ = 78 K.[10] Moreover, one can also replace S by Se while keeping the same crystal structure, adding more flexibility to the choice of materials. This diversity of different physical properties in these materials will turn out to be extremely useful from a materials' point of view when it comes to actual applications. More importantly, successful exfoliation has been recently demonstrated down to monolayer of these newly discovered vdW materials.[11-13]

With the successful initial works, it now needs further in-depth studies of its physical properties. So far all the initial characterization has been made by using Raman and AFM, leaving many of the physical properties literally unexplored. For example, there has been no report of the transport characteristics of few-layer thick flakes of $TMPX_3$, let alone monolayer although it is an essential step for future device applications. To measure current through ultrathin nanostructures such as thin oxide layers[14,15], molecules[16], and vdW materials[17,18], conductive atomic force microscopy (C-AFM) has become an instrument of choice. The imaging mode and the spectroscopic mode of C-AFM have been proven extremely useful for studying electronic properties on nanoscale. Especially, C-AFM is now an invaluable tool for mechanically exfoliated materials due to the small sample size of often microns or smaller. In this paper, we report on the tunneling transport properties of several exfoliated $MnPS_3$ flakes with thickness varying from monolayer to a few tens of layers by the spectroscopic mode of C-AFM. Using these data, we estimated several key parameters of $MnPS_3$ down to monolayer such as a tunneling barrier height and a dielectric breakdown strength.

In order to optimize the optical contrast between mono-layer $MnPS_3$ and Si substrate, we deposited 70 nm of transparent conductive indium tin oxide (ITO) on top of Si substrates: from our simulation we found this 70 nm thickness to give the maximum contrast between the substrate and the sample. All our single crystal samples of $MnPS_3$ were synthesized by a chemical vapor transport method: we put a quartz ampoule containing the raw materials into a horizontal 2-zone furnace with a temperature difference of 780 ℃ (hot zone) and 730 ℃ (cold zone). Details of the synthesis conditions can be found in literature [11]. Using these crystals and scotch tape, we exfoliated several samples with varying thickness from mono to a few tens of layers onto the substrate. We then carefully screened different parts of the samples by optical microscope and the thickness is confirmed by topological scans as shown in Figures 1(a) and 1(b). We found that thin $MnPS_3$ flakes are stable in air and no sign of degradations has been observed by optical and AFM images for a week when kept inside a desiccator after exfoliation. Topological and current scans were performed by C-AFM (Asylum Research Cypher S ORCA Conductive AFM) with a Pt-Ir coated Si tip (Nanosensors PPP-NCHPt). During the I-V measurement of the ITO/MnPS$_3$/Pt-Ir tip junction, bias voltage was applied to the substrate with the tip kept grounded. All measurements were performed within 4 hours after cleaving the samples.

We show the typical optical and topological images of mono- to 6 layers MnPS$_3$ in Figures 1(a) and 2(b). As one can see in the figures, atomically flat MnPS$_3$ was successfully isolated and the flake size is found to be bigger than 1 μm. The optical image was subsequently obtained using a closed aperture stop to enhance the contrast and each individual flakes are clearly distinguishable from one another in the images. The profile line in Figure 2(b) also shows that the thickness changes along the dashed line with the numbers representing the estimated thicknesses (nm) of mono-, bi-, and tri-layer flakes. Layer numbers were then determined by using the interlayer distance (0.65 nm) of MnPS$_3$,[19] and they are labeled accordingly in Figure 1(c). We note that the measured thickness is slightly larger than the expected value probably because of a thin water layer or a difference in the interaction force of tip with the substrate and MnPS$_3$ flake as discussed in Ref. [11]. A similar observation has been often made in other mechanically exfoliated vdW materials like graphene,[1,20] TMDCs,[21-23] and NiPS$_3$.[11,13]

It is rather common to observe carrier transport in a metal/insulator/metal junction being affected by a band alignment in accordance with the Schottky-Mott theory.[24] Figure 1(d) shows the expected band diagram of an ITO/MnPS$_3$/Pt-Ir tip in an equilibrium state suitable for our experimental set-up. In this schematic, positive bias voltage moves electrons from the Pt-Ir tip to the ITO-coated substrate and the electrons are supposed to go through the tunneling barrier. Therefore, we can determine the barrier height $\Phi_B$ from a difference between an electron affinity $\chi_s$ of MnPS$_3$ and a work function of Pt-Ir (5.4 eV). Due to the large barrier height at the surface of MnPS$_3$, we expect that the thick flakes are almost insulating as shown in Figure 1(c). In fact, our data show that even mono-layer MnPS$_3$ shows a very high resistance value while the ITO-coated substrate shows good conductivity during the fast current scan (1 Hz, 256 lines).

In Figure 2(a) we present some representative I-V spectroscopic data with various layer numbers of MnPS$_3$ flakes, which exhibits the typical Schottky junction behavior. For example, there is clear linear behavior at a low-bias voltage region while the current increases rapidly above the breakdown voltage. We comment that small asymmetric I-V characteristics observed in Figure 2(b) probably arises from the asymmetric band alignment of the junction in Figure 1(d). The breakdown voltages in the positive bias are little larger than that in the negative bias region because the barrier height of the tip is higher than that for the ITO. We also show in Figure 2(c) the dielectric breakdown voltage as a function of thickness, at which the current reaches a typical value of 1 nA.[25] From the slope in the linear fitting, we can determine the dielectric breakdown strength of thin MnPS$_3$ flake as 5.41 MV/cm. This value is smaller than that of SiO$_2$ (8-10 MV/cm)[26] and h-BN (7.89 MV/cm)[17], but it is a large enough value to make MnPS$_3$ a good insulating vdW material down to monolayer.

We also display the I-V characteristics for mono-, bi-, and tri-layer MnPS$_3$ in Figure 2(b). In this figure, the linear behavior is due to a direct tunneling at low bias voltage. And the change in the slope with layer numbers indicates the layer-number dependence of tunneling conductance: which decreases exponentially as the thickness increases as shown in inset of Figure 2(b). At low bias voltage (V), the tunneling current (I) can be expressed as follows:[17,27]

$$I(V) = \frac{A_{eff}\sqrt{m_e \phi_B} q^2 V}{h^2 d} \exp\left[\frac{-4\pi\sqrt{m_e \phi_B} d}{h}\right],$$

(1)

where $A_{eff}$ and d are an effective contact area and thickness of MnPS$_3$, respectively: q, $m_e$, and h are electron charge, free electron mass, and Plank constant, respectively. Thus, Eq. (1) naturally explains the exponential decay of the tunneling conductance as a function of d. According to our experiment, this direct tunneling is almost suppressed for samples thicker than 6 layers.

On the other hand, at high bias voltage the I-V curve does no longer show a linear dependency. Instead, as the tunneling barrier is deformed by the applied high electric field, the field-emitted electrons flow across the sample and the current becomes a non-linear function of applied voltage. This behavior can be understood by a Fowler-Nordheim tunneling (FNT) theory. In this case, the current can be expressed in the following formula:[14-18,28]

$$I(V) = \frac{A_{eff}q^3 m_e V^2}{8\pi h \phi_B d^2 m^*} \exp\left[\frac{-8\pi\sqrt{2m^*}\phi_B^{3/2} d}{3hqV}\right],$$

(2)

where $m^*$ is the effective electron mass for MnPS$_3$. Because of the lack of experimental and theoretical results for the effective mass of MnPS$_3$, we used $m^* = 0.5$ $m_e$ for our analysis to be followed below. For comparison, a similar value of the effective mass (0.56 $m_e$) was theoretically found for MnPSe$_3$[29] and the effective mass of MoS$_2$ is also known to be 0.35 $m_e$ for mono-layer and 0.53 $m_e$ for bulk.[18] Given the same crystal structure for both compounds, we anticipate that the real value of the effective mass of MnPS$_3$ may not be that different from MnPSe$_3$. With this assumption, Eq. (2) can be re-arranged as follows:

$$\ln\frac{I(V)}{V^2} = \ln\frac{A_{eff}q^3 m_e}{8\pi h \phi_B d^2 m^*} - \frac{8\pi\sqrt{2m^*}\phi_B^{3/2} d}{3hqV}.$$

(3)

Using this equation, we show the plot of $\ln(I(V)/V^2)$ versus $1/V$ from mono- to 9 layers MnPS$_3$ in Figure 3(a). Linear dependency observed in this plot indicates that the tunneling current at high bias voltage can be well explained by the FNT theory as expected.

Using the measured thickness of the sample and the assumed effective mass, we can calculate the effective contact area and the tunneling barrier height from the linear fitting shown in Figure 3(a). To get an independent verification of this approach and the values obtained, we have also separately carried out a direct FNT fitting by using Eq. (2) with the raw I-V data as shown in Figure 3(b). A good agreement between the two approaches gives us further confidence in the analysis we adopted here (see inset of Fig. 3(c)). The thus calculated barrier heights between the Pt-Ir tip and MnPS$_3$ are plotted as a function of thickness in Figure 3(c). From this curve, we obtain the barrier height value of 1.31 eV (±0.01) for samples thicker than 9 layers, which gets decreased with decreasing the layer number. This value is larger than that between Pt-Ir tip and MoS$_2$ (0.64 eV)[18] and smaller than that between Pt tip and h-BN (3.07 eV).[17] Our estimate of electron affinity is 4.09 eV for MnPS$_3$. We note that a similar reduction of the barrier height was previously reported for MoS$_2$[18], which may well come from the contribution of the direct tunneling in few layer thick samples. We further remark that a small mismatch was found in Figure 3(b) between the fitting curve and the experimental data just before the breakdown voltage. We think that it is most likely to come from the direct tunneling that decreases as the layer number gets increased. Therefore, the FNT model must overestimate the current for our thinner samples, and this results in the underestimated barrier height. The inset of Figure 3(c) shows the linear dependence of tunneling barrier term $\Phi_B^{3/2}d$ with a number of layers. From this analysis, we get an averaged effective contact area from mono- to 6 layers around 3.76 nm$^2$, which is a reasonable value considering the actual tip radius of 7 nm.[18]

In summary, we report the tunneling transport of mono- and few layers MnPS$_3$ exfoliated on top of ITO coated Si substrate. By using C-AFM, we could study the thickness dependence of the tunneling transport from a few tens of layers down to monolayer. From the I-V spectroscopy, we observed the Schottky junction tunneling behavior for our ITO/MnPS$_3$/Pt-Ir junction with a dielectric breakdown strength of 5.41 MV/cm, suggesting that an insulating behavior remains solid down to a monolayer of MnPS$_3$. At the same time, direct tunneling was suppressed at low bias voltage for MnPS$_3$ thicker than 4 layers. Using the FNT theory and assuming effective mass of 0.5 $m_e$, we estimate the barrier height of thin MnPS$_3$ flakes to be 1.31 eV (±0.01). We expect these values of tunneling barrier and dielectric breakdown strength would be helpful for many applications such as field effect transistor and magnetic tunnel junction.

The work at the IBS CCES was supported by the research program of Institute for Basic Science (IBS-R009-G1). S.L. and B.H.P were supported by the National Research Foundation of Korea (NRF) grants funded by the Korea government (MSIP) (No. 2013R1A3A2042120).

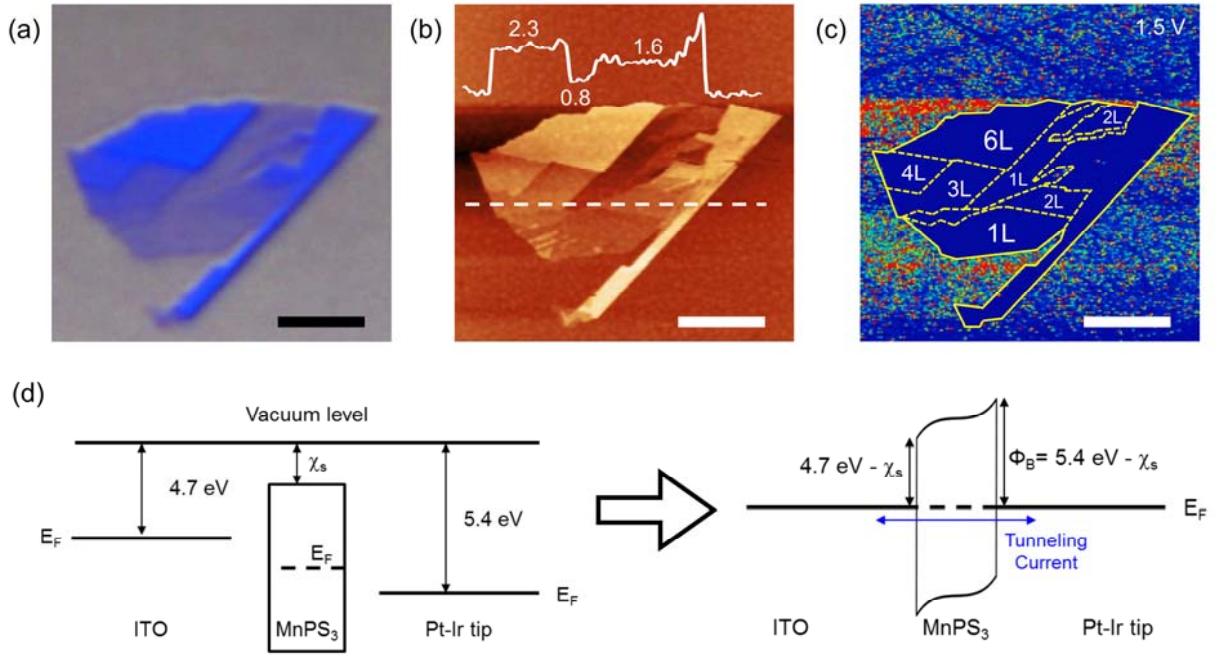

FIG. 1. (Color online) (a) Optical microscopic, (b) topological, and (c) current images of the mono-layer to 6 layers $MnPS_3$. The scale bars at the bottom of the images is for 5 μm. The profile line shows changes in the thickness (nm) along the dashed line with the numbers in nm on the top of the image (b). The current image (c) was obtained with a tip bias voltage of 1.5 V. (d) Schematic band alignment of the ITO/$MnPS_3$/Pt-Ir tip junction is shown in an equilibrium state appropriate for our experimental set-up. The barrier height $\Phi_B$ is estimated as the difference between the electron affinity $\chi_s$ of $MnPS_3$ and the work function of the Pt-Ir tip by the Schottky-Mott theory.[24]

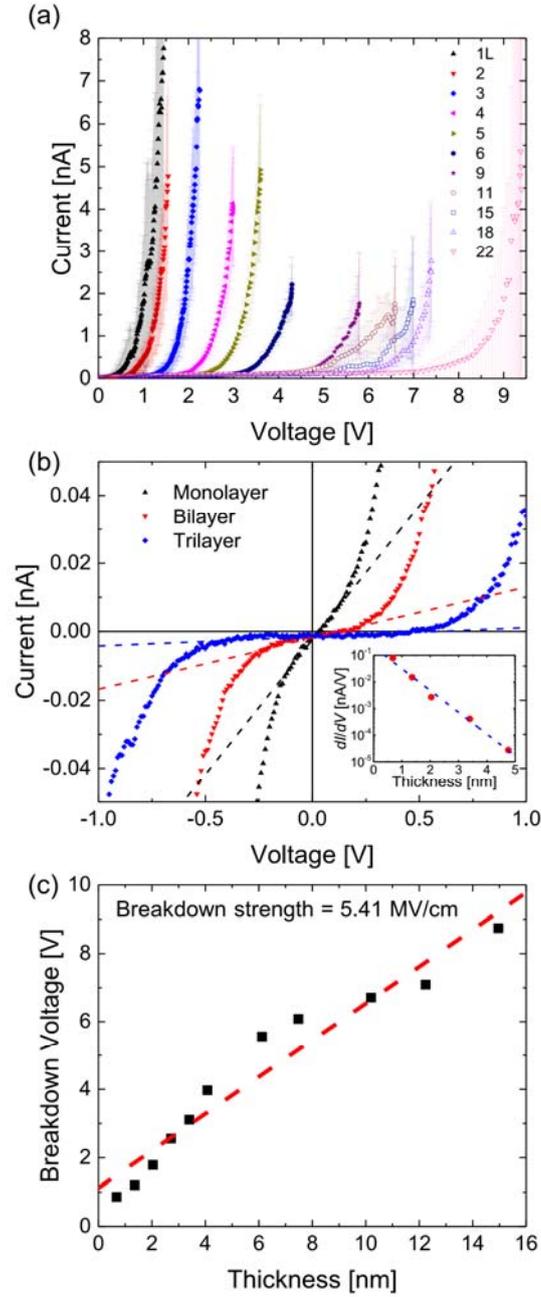

FIG. 2. (Color online) (a) I-V characteristics of ITO/MnPS$_3$/Pt-Ir tip junction for various thicknesses of MnPS$_3$. (b) Blown-up picture of I-V data at a low voltage region of mono-, bi-, and tri-layer MnPS$_3$. As voltage increases, a transition occurs from a direct tunneling to a FNT type. Inset of (b) indicates that the slope of the linear behavior at low voltage decreases exponentially as the number of layer gets increased. (c) Thickness dependence of breakdown voltage. The linear fit (dashed line) shows that the dielectric breakdown strength is 5.41 MV/cm for thin MnPS$_3$.

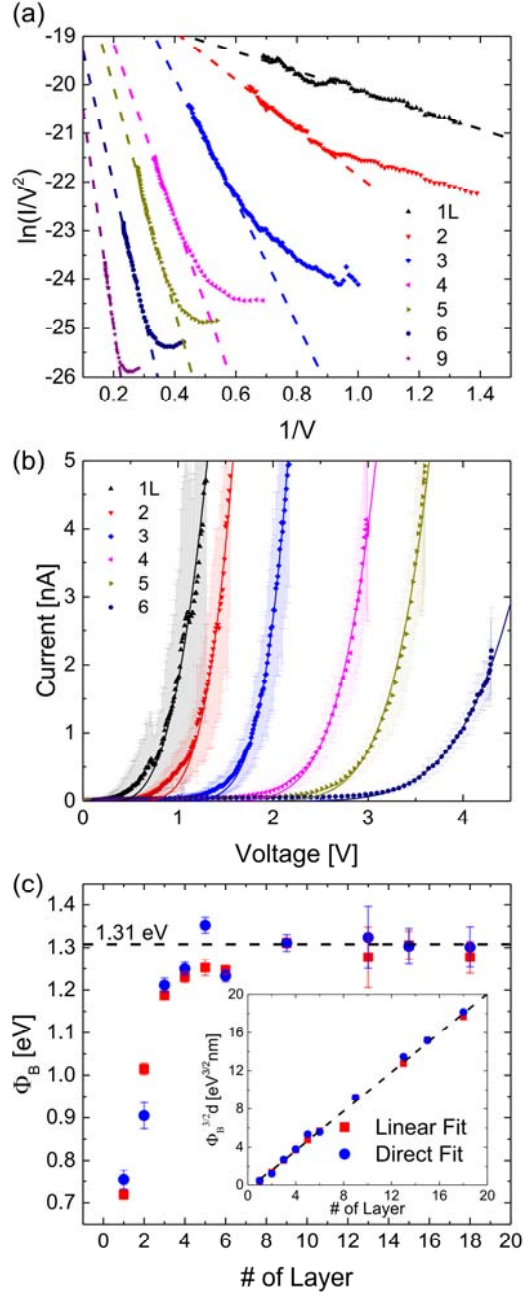

FIG. 3. (Color online) (a) $\ln(I/V^2)$ versus $1/V$ plots and their linear fit using the FNT model. The effective mass of thin MnPS$_3$ was assumed as $m^* = 0.5\, m_e$ for the fitting as discussed in the text. (b) Direct fitting (solid line) of the FNT model with the experimental I-V data. A slight deviation from the fitting at low voltage is most likely to be due to the direct tunneling and it gets suppressed as the thickness increases. (c) Calculated barrier height from both linear fit and direct fit. The barrier height was calculated as 1.31 eV (±0.01) for samples thicker than 9 layers, whereas for samples thinner than 6 layers the barrier height decreases as the number of layer decreases. Inset of (c) shows linear dependency of tunneling barrier term ($\Phi_B^{3/2}d$) on the number of layer.